\documentclass[]{aa}

\usepackage{graphicx}
\usepackage{txfonts}
\begin{document}
 
 \title{MUSE-AO view of the starburst--AGN connection: NGC~7130\thanks{Based on observations made at the European Southern Observatory using the Very Large Telescope under programme 60.A-9493(A).}}
 
 \author{J.~H.~Knapen\inst{1,2,3}, S.~Comer\'on\inst{4}, and M.~K.~Seidel\inst{5}}
         
    \institute{Instituto de Astrof\'isica de Canarias E-38205, La Laguna, Tenerife, Spain
              \and Departamento de Astrof\'isica, Universidad de La Laguna, E-38205, La Laguna, Tenerife, Spain
              \and Astrophysics Research Institute, Liverpool John Moores University, IC2, Liverpool Science Park, 146 Brownlow Hill, Liverpool, L3 5RF, UK
              \and University of Oulu, Astronomy Research Unit, P.O.~Box 3000, FI-90014 Oulu, Finland
              \and Caltech-IPAC, MC 314-6, 1200 E California Blvd, Pasadena, CA 91125, USA\\
              }
              
  \abstract{We present the discovery of a small kinematically decoupled core of  $0\farcs2$ (60\,pc) in radius as well as an outflow jet in the archetypical AGN--starburst "composite" galaxy NGC~7130 from integral field data obtained with the adaptive optics-assisted MUSE-NFM instrument on the VLT. Correcting the already good natural seeing at the time of our science verification observations with the four-laser GALACSI AO system, we reach an unprecedented spatial resolution at optical wavelengths of around $0\farcs15$. We confirm the existence of star-forming knots arranged in a ring of  $0\farcs58$ (185\,pc) in radius around the nucleus, previously observed from UV and optical {\it Hubble Space Telescope} and CO(6-5) ALMA imaging. We determine the position of the nucleus as the location of a peak in gas velocity dispersion. A plume of material extends towards the NE from the nucleus until at least the edge of our field of view at 2$^{\prime\prime}$ (640\,pc) radius which we interpret as an outflow jet originating in the AGN. The plume is not visible morphologically, but is clearly characterised in our data by emission-line ratios characteristic of AGN emission, enhanced gas velocity dispersion, and distinct non-circular gas velocities. Its orientation is roughly perpendicular to the line of nodes of the rotating host galaxy disc. A circumnuclear area of positive and negative velocities of  $0\farcs2$ in radius indicates a tiny inner disc, which can only be seen after combining the integral field spectroscopic capabilities of MUSE with adaptive optics.}


   \maketitle            

\section{Introduction}   

The accretion activity of supermassive black holes (SMBHs) is known to be closely related to the evolution of galaxies, and the mass of the SMBH scales with key galaxy parameters such as its velocity dispersion or the stellar mass and the luminosity of the bulge \citep[see review by][]{Kormendy2013}. These relationships may result from feedback from accretion onto a SMBH conditioning the evolution of the host galaxy, and/or from increased cold gas availability, which may simultaneously increase the star formation (SF) and the active galactic nucleus (AGN) activity.

A well-documented complication in the fuelling process of both AGN and circumnuclear starbursts is that the cold gaseous fuel, while plentiful in the body of the host galaxy, must lose by far most of its angular momentum before it can reach the central starburst and/or AGN \citep[e.g.,][]{Begelman1984, Shlosman1989}. Once it does, for instance under the influence of non-axisymmetries in the host galaxy induced by bars or past or present interactions, the availability of gaseous fuel can lead to both starburst and AGN activity, which has been referred to as ``composite'' if occurring simultaneously. Studying the gas physics, stellar properties, and kinematics of such composite AGN/starburst galaxies in detail can therefore provide fundamental clues on gas transport in, and the evolution of, galaxies, as well as the physics of AGN and circumnuclear starbursts.

One of the best-known examples of such composite AGN-starburst galaxies is the luminous infrared galaxy (LIRG) NGC~7130 (also known as IC~5135), a peculiar Sa \citep{deVaucouleurs1991} galaxy at a distance of 65.5\,Mpc (so $1^{\prime\prime}$ corresponds to 318\,pc). NGC~7130 hosts a Seyfert 1.9 AGN nucleus \citep{Veron-Cetty2006} as well as a powerful compact circumnuclear starburst \citep{Phillips1983}.


In this Letter, we present the first results from the science verification phase of MUSE-NFM (narrow field mode), which uses the adaptive optics (AO) system GALACSI with four laser guide stars to feed MUSE. 


\section{Observations and data reduction}

\label{observations}

\subsection{MUSE NFM AO}

\label{obs-muse}

\begin{figure*}[htb]
   \centering
   \includegraphics[width=1\textwidth]{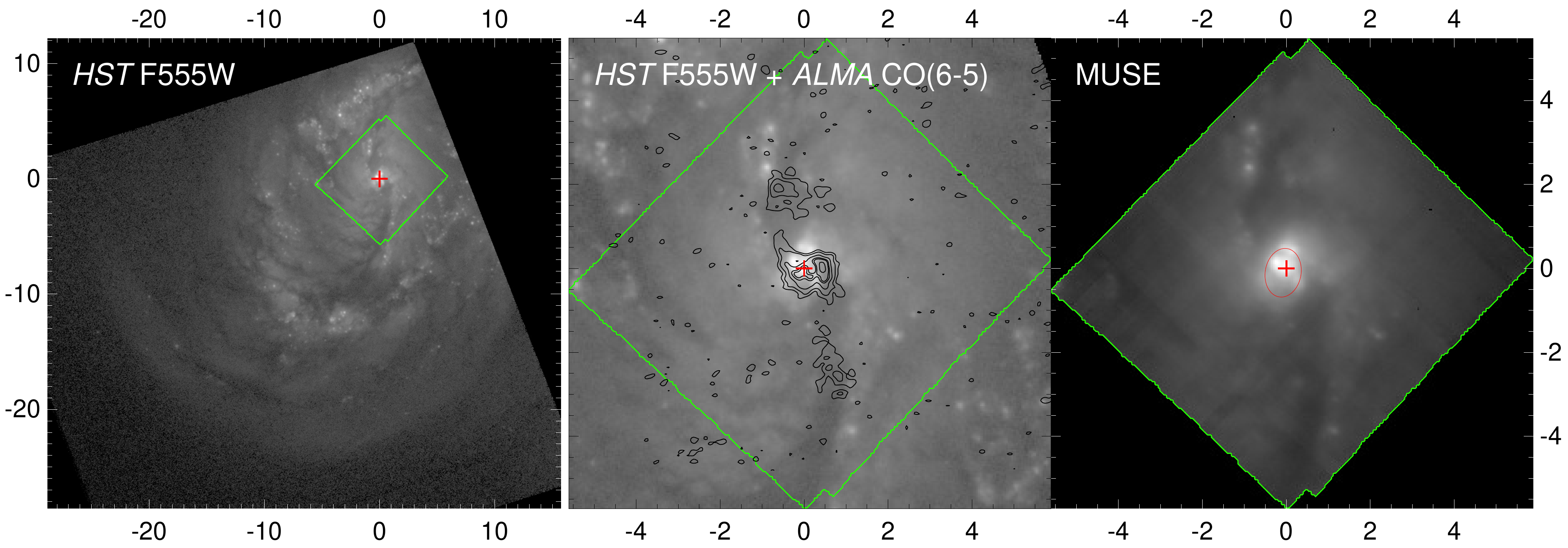}
   \caption{\label{fig1}{\it Left panel}: F555W {\it HST} image covering the central area of NGC~7130. Overlaid in green is the area covered by  MUSE-NFM. {\it Middle panel}: Zoomed-in view of the {\it HST} F555W image. Black contours correspond to the CO(6-5) line emission calculated as in \citet{Zhao2016} from their ALMA datacube; levels are as in \citet[][$\lbrack1, 2, 4, 6, 7, 10\rbrack\times 3\sigma$ with $\sigma=1.1\,{\rm Jy\,beam^{-1}\,km\,s^{-1}}$]{Zhao2016} but look slightly different, possibly because of additional processing performed by \citet{Zhao2016}. {\it Right panel}: MUSE datacube integrated along the spectral direction. We note the high spatial resolution of this image, similar to that in the {\it HST} image. The red ellipse indicates the UCNR, measured as in \citet{Comeron2014}. Coordinate labels are in arcseconds in all panels. The centre is defined to be at the spaxel with the largest sigma in the gas kinematics fit and is indicated by a red cross. North is up and east is to the left.}
\end{figure*}

We used the MUSE integral field spectrograph on the VLT in its newly commissioned NFM in conjunction with GALACSI, an AO system developed to increase the performance of MUSE. GALACSI employs four sodium laser guide stars, the deformable secondary mirror on the VLT's UT4, and an infrared low-order sensor to provide a near-diffraction-limited resolution at visible wavelengths (i.e., Strehl ratio>5\% at 650\,nm for $0\farcs6$ seeing). In addition to the four laser guide stars, we used the Seyfert nucleus of NGC~7130 as a natural guide star to measure the remaining atmospheric tip-tilt, defocus, and astigmatisms. Further technical details of MUSE with AO are given by \citet{Stuik2006}.

In its NFM, MUSE covers a field of view (FOV) of $7\farcs59$ by $7\farcs59$ with a sampling of $0\farcs0253$ per pixel. Our observations were taken for proposal number 60.A-9493(A) (PI M.~Seidel) during the nights of 2018 Sept 15, 16, and 18. The natural seeing on all three nights was below $1^{\prime\prime}$, but it was below $0\farcs6$ only on Sept~18. Only on that final night did the AO system deliver well enough; the spatial resolution from all other observations is poorer than $0\farcs5$ compared to the excellent $0\farcs15-0\farcs18$ on Sept~18. We obtained a total of ten sets of 600-s exposures on-source interlaced with 180-s exposures of offset sky, but only two sets during the good-seeing period on Sept~18. 

The resulting data were reduced using v.\,2.5.2 of the MUSE pipeline \citep{Weilbacher2012} under the ESO REFLEX interface \citep{Freudling2013} with the default parameters\footnote{We have been advised that the data reduction pipeline will be updated in early 2019. This update may affect the absolute flux calibration, so we warn that the values given here may have to be updated later. No other parameters are expected to be affected, and all other results presented here, in particular those based on line ratios and kinematics, are robust.}. The resulting datacubes were then manually aligned and the shifts were fed into \texttt{muse\_exp\_combine} to create a final combined datacube. We only combined the two datacubes with spatial resolution below $0\farcs2$.

To extract information from the spectra we followed the procedure in Comer\'on et al.~(2019, submitted), based on that used for MaNGA (M.~Cappellari, priv. comm.). To extract the data we considered only spaxels with ${\rm S/N}>0.5$ and produced two spatial Voronoi binnings using the software by \citet{Cappellari2003}: a {\it stellar tesselation} (ST) with a signal-to-noise ratio ${\rm S/N}\approx25$ in the stellar continuum between $5490$\,\AA\ and $5510$\,\AA\ (275\,bins), and an {\it emission line tesselation} (ET), with ${\rm S/N}\approx50$ for the H$\alpha$ line, obtained after subtracting the continuum using a spectral window at restframe $\lambda_{\rm H\alpha}+50$\,\AA\ (4645\,bins). 

We fitted the stellar component in the spectra obtained from the ST using \texttt{pPXF} \citep{Cappellari2004, Cappellari2017}. The code performs a full spectral fitting of the spectra with a linear combination of template spectral energy distributions widened to match the line of sight velocity distribution (LOSVD) of the spectral lines. We used the MIUSCAT template library \citep{Vazdekis2012}. As the spectral resolution of MUSE is worse than that of MIUSCAT in the blue, we convolved the MIUSCAT library with an adaptive Gaussian kernel to match the MUSE blue resolution. In red, the MUSE spectra were Gauss-convolved to match the MIUSCAT resolution. For this fit we masked the regions affected by strong sky lines and by emission lines. The fit was run in the range from 4750\,\AA\ to 8800\,\AA. The stellar continuum was modelled with an eighth-order additive Legendre polynomial.

We then reran \texttt{pPXF} over the spectra from the ST. This time, we kept the kinematics obtained in the previous step fixed and we did not mask the ISM emission lines. These were modelled with a Gaussian profile with the velocity and the velocity dispersion as free parameters. The lines fitted are H$\beta$, [O{\sc iii}]\,$\lambda4959$, [O{\sc iii}]\,$\lambda5007$, [O{\sc i}]\,$\lambda6300$, [O{\sc i}]\,$\lambda6364$, [N{\sc ii}]\,$\lambda6548$, H$\alpha$, [N{\sc ii}]\,$\lambda6583$, [S{\sc ii}]\,$\lambda6716$, and [S{\sc ii}]\,$\lambda6731$. All the fitted emission lines were tied to have the same velocity and velocity dispersion. We adopted as radial velocity of the galaxy the value of 4799\,km\,s$^{-1}$ from Zhao et al. (2016). 

We ran \texttt{pPPXF} a final time using the ET. Instead of fitting the stellar continuum with a combination of templates, we used the fitted stellar spectrum of the ST bin whose centre was the closest to that of the ET under study. The stellar model was not allowed to vary except for a scaling factor. This final \texttt{pPXF} fit is the one that provided the line properties reported in this Letter.

\subsection{{\it HST} and ALMA data}

\label{obs-rest}

We use additional data from the literature: an archival WFPC3 F555W {\it HST} image, 600\,s deep (6\,different exposures) and with $\sim0\farcs1$ angular resolution, taken as post-explosion data of SN~2010bt \citep{Elias-Rosa2018}, and the ALMA CO(6-5) data with a $0\farcs20\times0\farcs14$ resolution from \cite{Zhao2016}. The alignment of the various data sets is not trivial. We calibrated the astrometry for the {\it HST} and the MUSE data by comparing the position of SF knots with their coordinates in the Gaia DR2 \citep{Gaia2016, Gaia2018}. The ALMA data have good astrometry, so no calibration was required for those.

\section{Results}

\label{results}

\begin{figure*}[ht]
   \centering
   \includegraphics[width=0.8\textwidth]{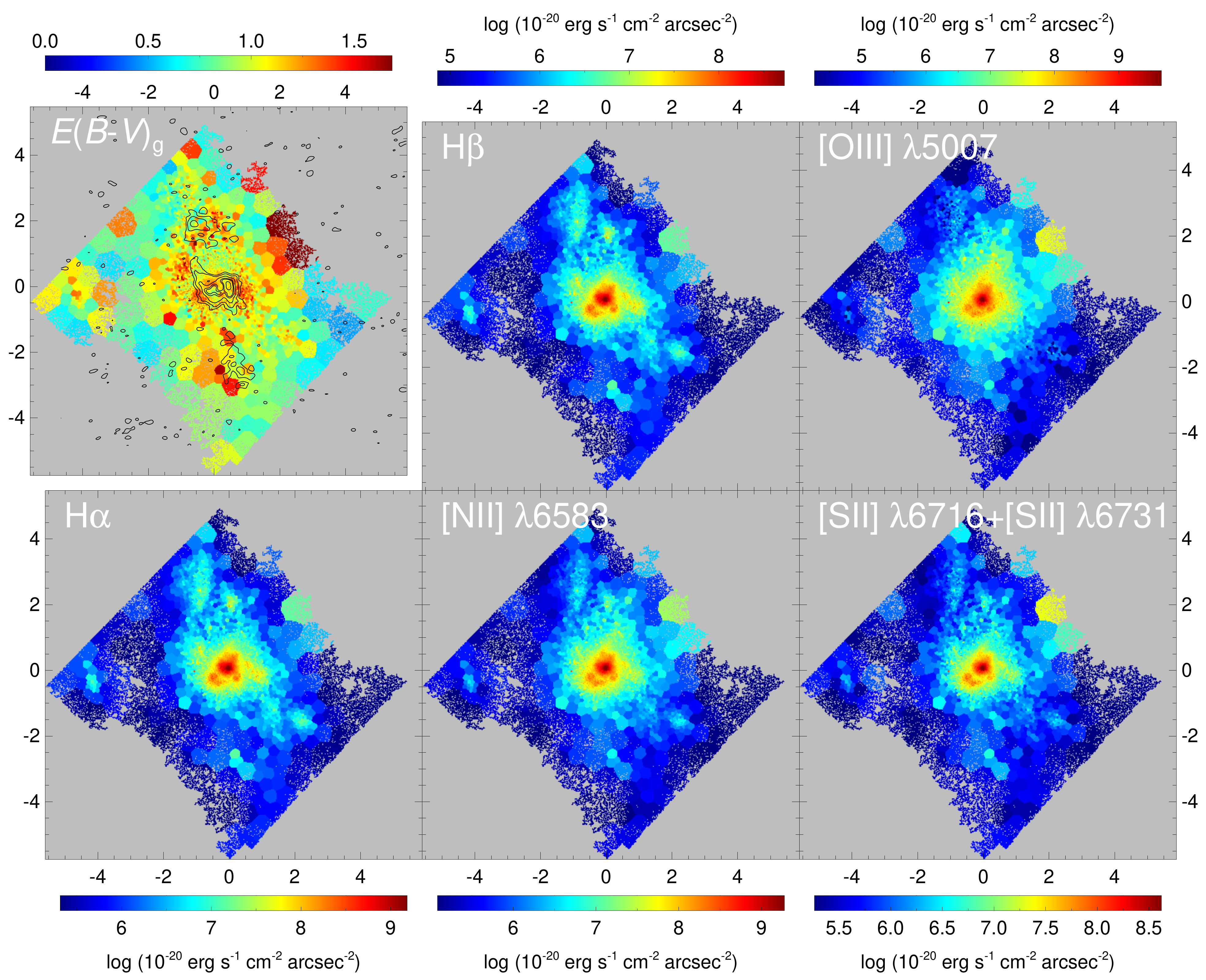}
   \caption{\label{fig2}{\it Top left panel}: $E(B-V)_{\rm g}$ dust extinction map derived from the H$\alpha$ and H$\beta$ MUSE maps. Overlaid are the CO(6-5) contours of Fig.~\ref{fig1}. {\it Other panels}: emission line maps in the lines of H$\beta$, [O{\sc iii}]\,$\lambda5007$, H$\alpha$, [N{\sc ii}]\,$\lambda6583$ and the [S{\sc ii}] doublet$^1$. Colour scale and units are indicated above and below each panel. Axis labels and orientation are as in Fig.~\ref{fig1}.}
\end{figure*}

The {\bf overall morphology} of the central region of NGC~7130 is well known, and the alignment of the CO(6-5) emission with darker patches indicating dust concentrations has been described by \citet{Zhao2016}. Figure~1 compares our MUSE data integrated over the whole spectral range  with archival {\it HST} and ALMA imaging, highlighting the structure of the central region with well-organised dust lanes, spiral arm fragments and bright emission from the central 1$^{\prime\prime}$, as well as the morphology of the CO and dust. Figure~\ref{fig1} shows the exquisite spatial resolution and imaging capacity of the new AO-assisted MUSE NFM: local peaks of emission can easily be recognised as having a very similar resolution to the {\it HST} image, and across most of the MUSE FOV. The spatial resolution of our MUSE data is around $0\farcs15$.

Various {\bf emission-line maps} as derived from our MUSE data are shown in Fig.~\ref{fig2}. These maps were corrected for extinction by assuming an unextincted ratio of $F_{\rm H\alpha}/F_{\rm H\beta}$ of $2.86$ \citep{Osterbrock1989}, and an extinction law as in \citet{Calzetti2000} with $R_V^{\prime}=4.05$. We show the resulting $E(B-V)_{\rm g}$ map in Fig.~\ref{fig2}.



The morphology of the central region is very similar in the different emission lines. In all maps, we see a strong nuclear peak, only slightly offset from the peak in velocity dispersion which we assume to be both the nucleus of the galaxy and the location of the AGN. This emission peak is surrounded, at a radius of $\sim0\farcs5$ ($\sim160\,{\rm pc}$), by the ring-like distribution of star forming regions first described by \citet{Gonzalez_Delgado1998} on the basis of their UV and optical {\it HST} imaging. Such small rings are not unusual, and were classified as ultra-compact nuclear rings (UCNRs) by \citet{Comeron2008}. In Fig.~\ref{fig2}, most of the individual fragments of emission in the {\it HST} UV map of \citet{Gonzalez_Delgado1998} can be seen, again confirming the superb quality of our new IFU data.

Outside the central arcsec region, various emission peaks can be identified in our maps, several of which confirm, but show much more detail of, the emission-line regions north and south of the nucleus identified by \citet{Davies2014}. The overall line emission follows a north-south distribution, highlighting SF occurring in the immediate vicinity of the dust lanes coming into the nuclear regions (see {\it HST} image, Fig.~\ref{fig1}). 

\begin{figure*}[h]
   \centering
   \includegraphics[width=0.8\textwidth]{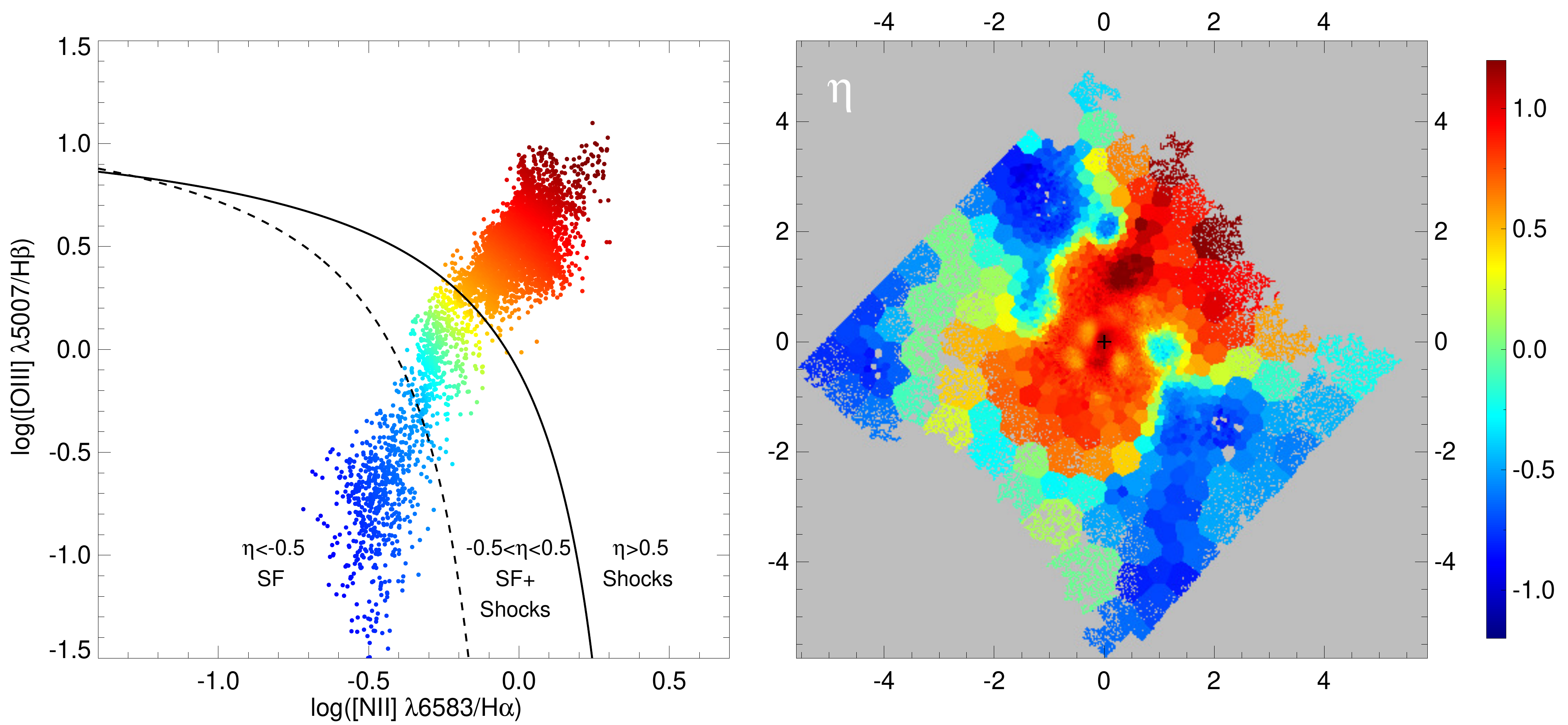}
   \caption{\label{fig3} O{\sc iii}/H$\beta$ vs N{\sc ii}/H$\alpha$ diagnostic diagram ({\it left}), populated with the individual spaxels from the MUSE cube and colour-coded by the $\eta$ parameter from Erroz-Ferrer et al.~(2019, submitted). The {\it right} panel indicates the location of the points according to these colours. The solid curve indicates the theoretical upper limit to pure SF from \citet{Kewley2001}, with the empirical limit to this from \citet{Kauffmann2003} given as the dashed curve. The cross denoting the centre and the axis labels and orientation in the right panel are as in Fig.~\ref{fig1}.}
\end{figure*}

A strong-line ratio {\bf diagnostic diagram} is shown in Fig.~\ref{fig3}. Such diagrams \citep[]{Baldwin1981} allow one to distinguish between H{\sc ii} region and AGN regimes of emission\footnote{To be more precise, between regions ionised by UV radiation from massive stars and those ionised by emission from AGN, shocks, or AGB stars}, and they are therefore excellent tools to study the details and origin of the starburst--AGN composite nature of the central region of NGC~7130. We colour-code the individual spaxels by their location in the diagnostic diagrams, and then map them back (Fig.~3, {\it right}) using the parameter $\eta$ as introduced by Erroz-Ferrer et al.~(2019, submitted) in the context of their analysis of the galaxies in the MUSE Atlas of Disks survey. These authors define $\eta$ as: {\it a) for the ``Shock region''} the orthogonal distance to the continuous curve, {\it b) for the ``SF'' region} the orthogonal distance to the dashed curve, and {\it c) for the region in between} the orthogonal distance to the bisector of the dashed and continuous curves. The parameter is normalised so that $\eta=0.5$ at the continuous curve and $\eta=-0.5$ at the dashed curve.



Figure~\ref{fig3} shows that there is a large spread of line ratios from the SF to the AGN parts of the diagrams, with the AGN-region spaxels located primarily in the nuclear region of the galaxy, surrounded by areas dominated by SF. This confirms the general findings of \citet{Davies2014}, but we do not confirm the almost perfectly annular distribution they present in their Fig.~5, indicating how the AGN contribution gradually decreases with increasing radius. Instead, we find a large plume of AGN-style emission extending from the nuclear region towards the NW, to the edge of our MUSE FOV, and with the largest values of $\eta$ occurring not in the nucleus but along the plume. Below, we come back to this plume, and propose that it traces an outflow from the AGN. The fact that the circumnuclear medium is affected by the AGN was also shown by \citet{Pozzi2017} from their study of the CO spectral energy distribution.

Another feature to note in Fig.~\ref{fig3} is that the location of the intensely star-forming regions \citep[e.g.,][and Figs.~\ref{fig1} and \ref{fig2}]{Gonzalez_Delgado1998} is nicely reflected in the right panel as distinct local spots with lower $\eta$ values in the central-arcsecond region in the top-right panel (seen as orange spots in the figure), the latter no doubt due to mixing of emission from the star-forming regions with that from the AGN.

\begin{figure*}[h]
   \centering
   \includegraphics[width=0.8\textwidth]{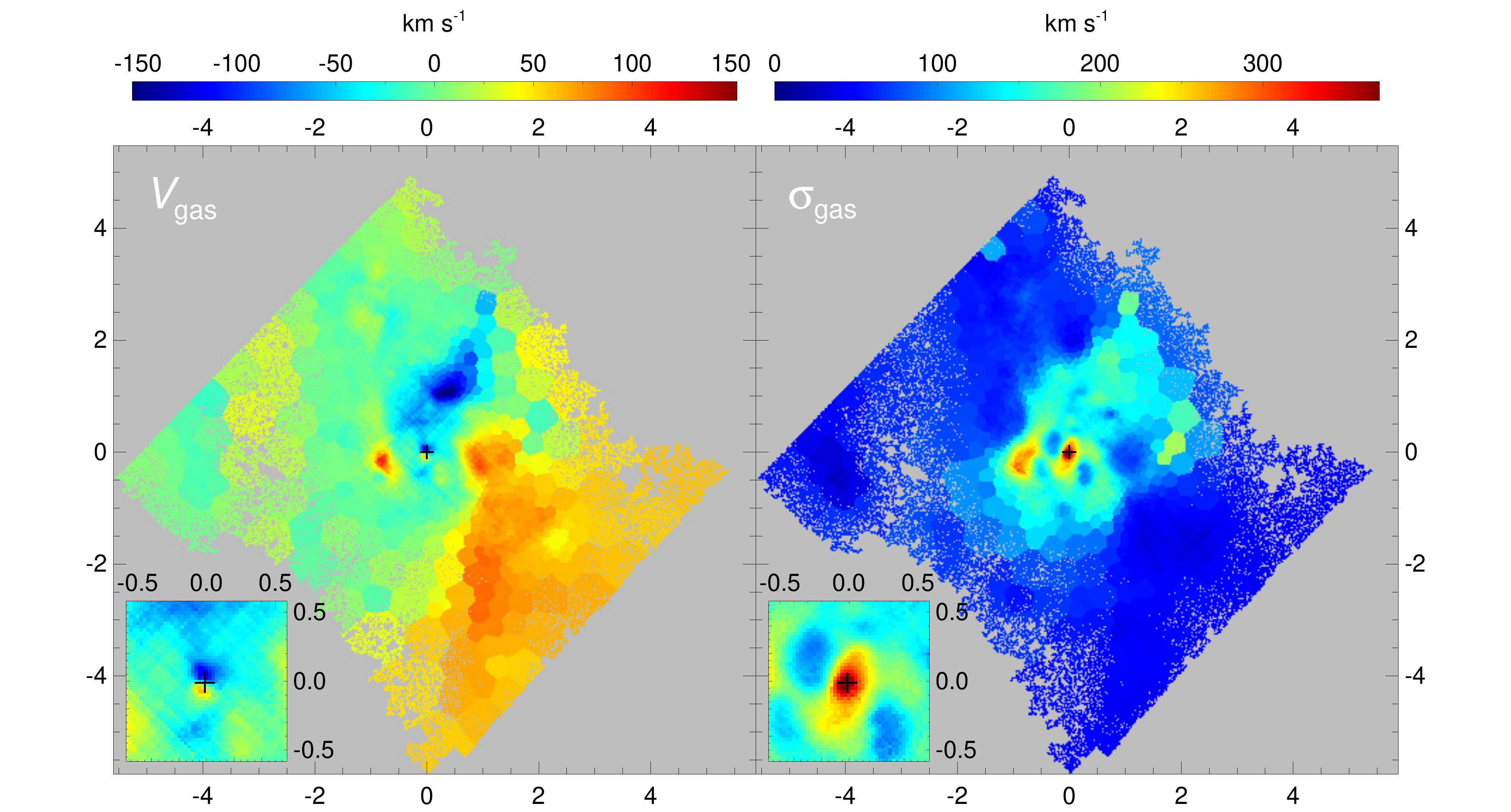}
   \caption{\label{fig4} Gas velocity field ({\it left panel}) and velocity dispersion map ({\it right panel}) derived from the ten emission lines (Sect.~\ref{observations}) in our MUSE data set. The insets show an enlarged version of the innermost $1\farcs2\times1\farcs2$. Colour scale and units are indicated above each panel. Following \citet{Zhao2016}, the zero velocity corresponds to a heliocentric recession velocity of $v_{\rm helio}=4799\,{\rm km\,s^{-1}}$. The crosses indicate the centre of coordinates. Axis labels and orientation are as in Fig.~\ref{fig1}.}
\end{figure*}

Figure~\ref{fig4} shows the {\bf emission line kinematics}: velocity field and velocity dispersion. From \citet{Zhao2016} we know that the gradient in the velocity field due to galaxy rotation is roughly east-west, and this is indeed seen in the area in Fig.~\ref{fig4}. We will explore this in more detail in a future paper, where we will combine the MUSE kinematics with H$\alpha$ Fabry-P\'erot data which we have just obtained at the 4.2\,m William Herschel Telescope. In that paper, we will also model the velocity field in more detail, using a multi-component analysis as in \citet{Davies2014} rather than the main-peak approach followed here in this initial work.

A tiny area of ordered rotation at enhanced velocities can be distinguished in the very central region (within $0\farcs2$ in radius, see inset in {\it left} panel in Fig.~\ref{fig4}). We interpret this as a tiny (radius 60\,pc) kinematically decoupled nuclear disc, with a rotation axis offset by around $90\degr$ from that of the main body of the galaxy. Such discs are not uncommon, but discovering such a small one around the Seyfert nucleus of NGC~7130 is proof of the capabilities of MUSE. It will be modelled in detail in a subsequent paper.

The velocity dispersion map shows a well-defined peak, which we identify as marking the location of the centre of the galaxy, and of the AGN. It is surrounded by an area of $\sim2^{\prime\prime}$ diameter with increased velocity dispersion and some localised enhanced regions which may be related to the star-forming spots in the UCNR. 

Another interesting feature in both the velocity and dispersion maps is the plume of approaching, high-dispersion gas extending from the nucleus to the edge of the MUSE FOV in the NW direction, coinciding spatially with the plume of material that we discovered from the diagnostic diagrams. 

\section{Discussion}

\label{discussion}

\subsection{Nuclear disc}

The exquisite spatial resolution of our new MUSE-NFM data set is highlighted by the recovery of almost all the structure in the circumnuclear region, which previously had only been seen in {\it HST} UV and optical imaging \citep[e.g.~][]{Gonzalez_Delgado1998}. We confirm the existence of star-forming regions in a ring configuration with a radius of $0\farcs58$ (185\,pc; outlined in Fig.~\ref{fig1}), a tiny ring that is classified as an UCNR following the size criterion in \citet{Comeron2008}, and whose size we measure here following the methodology of \citet{Comeron2014}. But what our MUSE data offer is a whole new angle on the physics of these circumnuclear regions, ranging from their emission line ratios (showing locally enhanced line ratios due to SF; Fig.~\ref{fig3}) and kinematics (localised streaming motions and variations in velocity dispersions, Fig.~\ref{fig4}) presented in this Letter, to their metallicity and stellar populations, which will be studied in future papers.

The small region with excess positive and negative velocities surrounding the peak in velocity dispersion and indicating the nucleus of the galaxy (inset, {\it left} panel, Fig.~\ref{fig4}) has its velocity field oriented in the same direction as the jet and may be connected to it, although an alternative interpretation as a tiny nuclear disc rotating in a different plane from the host galaxy would not be at odds with the jet explanation. 

\subsection{Outflow plume}

A second striking feature discovered from our new MUSE data is the plume of low-velocity, high-velocity dispersion material with AGN-like emission line characteristics extending from the nucleus towards the NW direction, and reaching the edge of the FOV, $2^{\prime\prime}$ or 640\,pc from the nucleus (Figs.~\ref{fig3}, \ref{fig4}). We interpret this plume as material flowing out from the Seyfert core in a one-sided jet configuration. Its orientation is almost exactly $90\degr$ offset from the line of nodes, which could indicate polar emission if the black hole accretion disc is rotating in the same plane and sense as its host galaxy. 
 
The plume towards the NW does not have a counterpart in either UV, optical, or IR imaging \citep{Gonzalez_Delgado1998, Davies2014}, nor in CO(6-5) \citep{Zhao2016}. Its presence can also not be detected from individual narrow-band images or the dust extinction map (Fig.~\ref{fig2}). It is only through sensitive high-resolution IFU data such as those from MUSE presented here that it can be uncovered, through its AGN-like emission line ratios, and through its kinematics. Its discovery adds another fascinating component to the rich morphology and physics of this composite AGN+starburst galaxy hosting one of the few known UCNRs.
 
\section{Conclusions}

\label{conclusions}

New IFU data obtained with the AO-assisted MUSE-NFM instrument on the VLT of the archetypical AGN--starburst ``composite'' galaxy NGC~7130, achieving an unprecedented spatial resolution at optical wavelengths of around $0\farcs15$, has led to two discoveries.

Firstly, we discover a distinct circumnuclear area of $0\farcs2$ in radius of positive and negative velocities which indicate a tiny kinematically decoupled inner disc. Such discs are commonly seen as a remnant of past interactions between galaxies, but the small size and proximity to the composite AGN+starburst core of NGC~7130 adds a new dimension to our understanding of this system.

Secondly, we report the discovery of a plume of material extending towards the NE from the nucleus until at least the edge of our FOV at $2^{\prime\prime}$ (640\,pc) radius which we interpret as an outflow jet originating in the AGN. The plume is not visible morphologically, but is clearly characterised in our data by emission-line ratios characteristic of AGN emission, enhanced gas velocity dispersion, and distinct non-circular gas velocities. Its orientation is roughly perpendicular to the line of nodes (Fig.~\ref{fig4}) of the rotating host galaxy disc, which indicates that the black hole powering the AGN may rotate in the same orientation. 

More detailed analysis and modelling of the physical properties and the kinematics of both the inner disc and the plume regions will be presented in future papers.

\begin{acknowledgements}
 
 We thank the staff who made MUSE perform so well in its new NFM mode. We thank Dr.~Lodovico Coccato for his help with the data reduction, Dr.~Ana Monreal-Ibero for comments on the manuscript, and Dr.~Fatemeh Tabatabaei for advice on the astrometric calibration. J.H.K.~acknowledges financial support from the European Union's Horizon 2020 research and innovation programme under Marie Sk{\l}odowska-Curie grant agreement No 721463 to the SUNDIAL ITN network, from the Spanish Ministry of Economy and Competitiveness (MINECO) under grant number AYA2016-76219-P, from the Fundaci\'on BBVA under its 2017 programme of assistance to scientific research groups, for the project ``Using machine-learning techniques to drag galaxies from the noise in deep imaging'', and from the Leverhulme Trust through the award of a Visiting Professorship at LJMU. This research has made use of the NASA/IPAC Extragalactic Database (NED), which is operated by the Jet Propulsion Laboratory, California Institute of Technology, under contract with the National Aeronautics and Space Administration. This Letter makes use of the following ALMA data: ADS/JAO.ALMA\#2013.1.00524.S. ALMA is a partnership of ESO (representing its member states), NSF (USA) and NINS (Japan), together with NRC (Canada), MOST and ASIAA (Taiwan), and KASI (Republic of Korea), in cooperation with the Republic of Chile. The Joint ALMA Observatory is operated by ESO, AUI/NRAO and NAOJ. This work has made use of data from the European Space Agency (ESA) mission {\it Gaia} (\url{https://www.cosmos.esa.int/gaia}), processed by the {\it Gaia} Data Processing and Analysis Consortium (DPAC, \url{https://www.cosmos.esa.int/web/gaia/dpac/consortium}). Funding for the DPAC has been provided by national institutions, in particular the institutions participating in the {\it Gaia} Multilateral Agreement. Some of the data presented in this Letter were obtained from the Mikulski Archive for Space Telescopes (MAST). STScI is operated by the Association of Universities for Research in Astronomy, Inc., under NASA contract NAS5-26555.

\end{acknowledgements}

\bibliographystyle{aa}
\bibliography{NGC7130}

\end{document}